 \documentclass[final,3p,times,twocolumn]{elsarticle}

\usepackage{amssymb}

\biboptions{sort&compress}

\journal{NIM B381(2016)16}

\begin{document}

\begin{frontmatter}



\title{Activation cross-sections of proton induced reactions on vanadium in the 37-65 MeV energy range}


\author[1]{F. Ditr\'oi\corref{*}}
\author[1]{F. T\'ark\'anyi}
\author[1]{S. Tak\'acs}
\author[2]{A. Hermanne}
\cortext[*]{Corresponding author: ditroi@atomki.hu}

\address[1]{Institute for Nuclear Research, Hungarian Academy of Sciences (ATOMKI),  Debrecen, Hungary}
\address[2]{Cyclotron Laboratory, Vrije Universiteit Brussel (VUB), Brussels, Belgium}

\begin{abstract}
Experimental excitation functions for proton induced reactions on natural vanadium in the 37-65 MeV energy range were measured with the activation method using a stacked foil irradiation technique. By using high resolution gamma spectrometry cross-section data for the production of $^{51,48}$Cr, $^{48}$V, $^{48,47,46,44m,44g,43}$Sc and $^{43,42}$K were determined. Comparisons with the earlier published data are presented and results predicted by different theoretical codes (EMPIRE and TALYS) are included. Thick target yields were calculated from a fit to our experimental excitation curves and compared with the earlier experimental yield data. Depth distribution curves to be used for thin layer activation (TLA) are also presented.
\end{abstract}


\end{frontmatter}


\section{Introduction}
\label{1}
Many authors published experimental cross-section data for proton induced reactions on natural vanadium, especially at lower energies in the frame of basic nuclear physics research \cite{1,2,3,4}.  A few works reported experimental thick target yields \cite{5,6,7,8}. As mentioned, most of the literature cross-sections data were obtained in the low energy region (below 30 MeV incident energy) and only a few (in some cases contradictory) data sets were measured at higher energy. We participated in an IAEA-CRP setting-up an activation data library for proton and deuteron induced reactions for the FENDL project. Vanadium was included into this library as an important construction and alloying material in accelerator technology. During compilation of the experimental data and comparison with the TENDL library it was recognized that the activation data at higher energies are contradictory and the theoretical model description for radionuclides far from the target mass number is very poor. The maximum proton energy presently available for us was 65 MeV at the Cyclone110 cyclotron of Louvain la Neuve (LLN) Belgium through the HAS-FWO (Vlaanderen) collaboration. We decided therefore to re-measure the cross-section data for all possible $\gamma$-emitting activation products of $^{nat}$V in the 37-65 MeV proton energy range. 
Natural vanadium consists of two isotopes: for more than 99.75 \% it is formed by stable $^{51}$V while the very long-lived $^{50}$V (T$_{1/2}$ = 1.4 1017 a) occurs only in 0.25 \%. Experimental data on $^{nat}$V can hence be interpreted in broad energy range as coming from a monoisotopic $^{51}$V target and reaction cross-sections can be derived.
Among the reaction products $^{51}$Cr is an attractive radionuclide for use in diagnostic nuclear medicine due to its favorable nuclear characteristics such as a half-life of 27.7010 days and decay by electron capture, resulting in emission of gamma photons with energy of 320.0824 keV (9.91 \%). Its main applications is labeling of red blood cells for measurement of mass or blood volume, labeling of platelets to determine their survival time as well as in sequestration studies for the diagnosis of gastrointestinal bleeding \cite{9}. The favorite routes for no carrier added production are the (p,n) and (d,2n) reactions on vanadium \cite{10, 11}.
Among the scandium radionuclides $^{44}$Sc, $^{47}$Sc and $^{46}$Sc have been used in biological and medical studies. The $^{47}$Sc has attracted attention because of its favorable decay characteristics (half-life: 3.35 d; average electron energy: 162 keV; E$_{\gamma}$: 159 keV) for therapeutic application and for SPECT imaging. The applicability of $^{44g}$Sc as a matching pair for $^{47}$Sc through PET imaging, $^{44m}$Sc for radionuclide therapy. The longer-lived $^{46}$Sc was used for regional blood flow studies and investigation of radiochemistry of scandium. Recently $^{46}$Sc has been advocated as an additional monitor for proton and deuteron reactions and will be taken up in the recommended data base of the IAEA-CRP \cite{12}. The main charged particle production routes to gain carrier free Sc radioproducts include proton and deuteron induced reactions on calcium, titanium and vanadium.
$^{43}$Sc was shown to be an excellent tool for studying blood perfusion in the heart and for diagnosing myocardial infarcts. The main production routes at charged particle accelerators are the $^{40}$Ar($\alpha$,p), $^{nat}$Ca(p,x) or $^{nat}$Ca(d,x) and high energy reactions on Ti and V, but  in these  last cases the cross sections are small.

\section{Experimental and data processing}
\label{2}
The excitation functions for the $^{nat}$V(p,x) reactions were measured via an activation method by using the stacked foil irradiation technique followed by high resolution $\gamma$-ray spectrometry without chemical separation. 
The irradiation was performed at the LLN Cyclone110 cyclotron with 65 MeV protons for 1 h at 35 nA beam intensity.  The stack contained a 19 times repeated set of 10 $\mu$m Al, 116 $\mu$m In, 99.2 $\mu$m Al, 8.41 $\mu$m V, 99.2 $\mu$m Al, 26.2 $\mu$m Ho and 99.2 $\mu$m Al foils. The Al-monitors were also used as recoil catchers as well as for exact determination of beam intensity and energy by re-measuring the excitation function for the $^{27}$Al(p,x)$^{22,24}$Na reactions over the entire energy domain \cite{13}. The target stacks were irradiated in a Faraday-cup like target holder, equipped with a long collimator (effective beam diameter on target is 5 mm). 
The gamma activity of the produced radionuclides was measured with standard high purity Ge detectors at the Cyclotron Laboratory of VUB-Brussels. 
Measurements of the induced activity started about 20 h after EOB (End of Bombardment), hence only activation products with a half-life longer than 3 h could be assessed. The samples were counted repeatedly at varying sample detector distance for weeks after EOB. The evaluation of gamma-spectra was made by automatic peak recognition programs and in a manually controlled interactive way.
For most of the radionuclides assessed different independent $\gamma$-lines are available, allowing an internal check of the consistency of the calculated activities.
The cross-sections were calculated by using the well-known activation formula with input parameters: measured activity, particle flux and number of target nuclei. In this study only a couple radionuclides formed as a result of a cumulative process by decay of a parent nuclide. The exact physical situation will be discussed individually for each activation product.
The decay and spectrometric characteristics, needed to transform count rates to activity of the different activation products at EOB, were taken from the NUDAT2 data base \cite{14}  and are summarized in Table 1. This table also includes the reaction Q-values of the contributing reactions \cite{15}.
The number of incident particles was initially derived from continuously measured beam current on target. The mean energy in each target foil was estimated by a stopping calculation from the incident beam energy and target thickness \cite{16}. 
The beam energy and intensity parameters were further adapted by taking into account the comparison of the excitation function of the $^{27}$Al(p,x)$^{22,24}$Na reactions, re-measured over the whole energy domain studied, with the recommended values in the updated version of IAEA-TECDOC 1211 \cite{17}. 
The uncertainties of the median energies were estimated from the uncertainties of the cumulative contributing processes (taking into account possible incident energy variation, thickness variation of the different targets and straggling effects) resulting in around $\pm$ 0.3 MeV for the first foil and about $\pm$ 1.3 MeV for the last foil.
The uncertainty on each cross-section point was estimated in the standard way \cite{18} taking the square root of the sum in quadrature of all individual contributions, supposing equal sensitivities for the different parameters appearing in the formula. The following individual uncertainties are included in the propagated error calculation: absolute abundance of the used $\gamma$-rays (5 \%), determination of the peak areas including statistical errors (4-10 \%), the number of target nuclei including non-uniformity (5 \%) and detector efficiency (10 \%). The total uncertainty of the cross-section values was estimated to be 8–14 \%. The strongly non-linear effect of the possible uncertainty of the half-lives, cooling time and measuring time for samples was not taken into account.

\begin{table*}[t]
\tiny
\caption{Decay characteristics of the investigated reaction products}
\centering
\begin{center}
\begin{tabular}{|p{0.4in}|p{0.4in}|p{0.4in}|p{0.4in}|p{0.3in}|p{0.6in}|p{0.4in}|} \hline 
Nuclide\newline Spin\newline Isomeric level & Half-life & Decay method\newline (\%) & E$_{\gamma}$(keV) & I$_{\gamma}$(\%) & Contributing process & Q-value\newline (keV) \\ \hline 
$^{51}$Cr \newline 7/2$^{-}$ \newline  & 27.7010 d & $\varepsilonup$: 100 & 320.0824  & 9.910 & $^{50}$V(p,$\gamma$)\newline $^{51}$V(p,n) & 9516.17\newline -1534.977 \\ \hline 
$^{48}$Cr\newline  0$^{+}$ & 21.56 h & $\varepsilonup$: 100  & 112.31\newline 308.24 & 96.0\newline 100 & $^{50}$V(p,3n)\newline $^{51}$V(p,4n) & -23327.18\newline -34378.33 \\ \hline 
$^{48}$V\newline 4$^{+}$~\newline  & 15.9735 d & $\varepsilonup$: 100 & 983.525\newline 1312.106 & 99.98\newline 98.2\newline  & $^{50}$V(p,p2n)\newline ${}^{51}$V(p,p3n)\newline $^{48}$Cr decay & -20889.72\newline -31940.86\newline -23327.18 \\ \hline 
$^{48}$Sc\newline 6$^{+}$ & 43.67 h & $\betaup^{-}$: 100 & ~175.361\newline 1037.522 & 7.48\newline 97.6 & $^{50}$V(p,3p)\newline $^{51}$V(p,3pn) & -19298.38\newline -30349.52 \\ \hline 
$^{47}$Sc\newline 7/2$^{-}$ & 3.3492 d & $\betaup^{-}$: 100 & ~159.381 & 68.3 & $^{50}$V(p,3pn)\newline $^{51}$V(p,3p2n) & -27537.48\newline -38588.63 \\ \hline 
$^{46}$Sc\newline 4$^{+}$ & 83.79 d & $\betaup^{-}$: 100 & ~889.277\newline 1120.545 & ~99.9840\newline 99.9870~ & ${}^{50}$V(p,3p2n)\newline ${}^{51}$V(p,3p3n) & -38183.93\newline -49235.08 \\ \hline 
$^{44m}$Sc\newline 6$^{+}$\newline 271.24 keV & ~58.61 h & $\varepsilonup$: 1.20\newline IT: 98.80 & 271.241 & 86.7 & $^{50}$V(p,3p4n)\newline $^{51}$V(p,3p5n) & -58271.09\newline -69322.24 \\ \hline 
$^{44g}$Sc\newline 2$^{+}$ & 3.97 h & $\varepsilonup$: 100 & 1157.020 & 99.9 & $^{50}$V(p,3p4n)\newline $^{51}$V(p,3p5n)\newline $^{44}$Ti decay & -58271.09\newline -69322.24\newline -59320.84 \\ \hline 
${}^{43}$Sc\newline 7/2$^{-}$ & 3.891 h & $\varepsilonup$: 100 & 372.9 & 22.5 & $^{50}$V(p,3p5n)\newline $^{51}$V(p,3p6n)\newline $^{43}$Ti decay & -67970.29\newline -79021.44\newline -75619.65 \\ \hline 
$^{43}$K\newline 3/2$^{+}$ & 22.3 h & $\betaup^{-}$: 100 & 372.760\newline 396.861\newline 593.390\newline 617.490 & 86.80\newline 11.85\newline 11.26\newline 79.2 & $^{50}$V(p,5p3n)\newline $^{51}$V(p,5p4n)\newline $^{51}$V(p,2$\alpha$p)\newline  & -66018.3\newline -77069.445\newline -20478.117 \\ \hline 
$^{42}$K\newline  & 12.360 h & $\betaup^{-}$: 100 & ~1524.6 & 18.08 & $^{50}$V(p,5p4n)\newline $^{51}$V(p,5p5n)\newline ${}^{51}$V(p,2$\alpha$pn) & -75642.984\newline -77069.445\newline -30102.797 \\ \hline 
\end{tabular}

\end{center}
\begin{flushleft}
\tiny{\noindent $\varepsilon$ is total EC + $\beta^+$
The Q-values refer to formation of the ground state. In case of formation of a higher laying isomeric state it should be corrected with the level energy of the isomeric state shown in Table 1. When complex particles are emitted instead of individual protons and neutrons the Q-values have to be decreased by the respective binding energies (pn $\longrightarrow$ d +2.2 MeV, p2n $\longrightarrow$ t +8.5 MeV, 2pn $\longrightarrow$ $^3$He +7.7 MeV, 2p2n $\longrightarrow$ $\alpha$ +28.3 MeV).
}
\end{flushleft}

\end{table*}




\section{Model calculations}
\label{3}
The cross-sections of the investigated reactions were compared with the data given in the last two on-line TENDL libraries to show the development of the predictions  (from TENDL-2010 to TENDL-2014 and TENDL-2015) \cite{19}. These libraries  are based on both default and adjusted TALYS calculations (Koning 2012)\cite{20}. For comparison also the results of the latest version of the EMPIRE code (EMPIRE 3.2 (Malta) are presented \cite{21, 22}. Default input was used in the most cases, expect the TENDL libraries, where the TALYS results are adjusted in an unknown (for the public user) way.

\section{Results}
\label{4}
The cross-sections for all the reactions studied are shown in Figs. 1–11 and the numerical values are collected in Table 2. No practical contribution from the reaction on the low abundance $^{50}$V can be distinguished, hence all results are practically equivalent to reaction cross sections on $^{51}$V.

\subsection{Production of $^{51}$Cr}
\label{4.1}
Many experimental data sets exist for production of $^{51}$Cr (T$_{1/2}$ = 27.701 d) especially at low energies for basic research.  Results were published by Tanaka, Albert, Shore, Taketani, Wing, Hansen, Albouy, Hontzeas, Humes, Johnson 1964, Dell, Harris, Chodil, Johnson 1958, Gadioli, Barandon, Mehta, Michel 1979, Michel 1980, Zyskind, Kailas, Bastos, Levkovskij(normalized), Jung, Musthafa and Carlson \cite{1,2,23,24,25,26,27,28,29,30,31,32,33,34,35,36,37,38,39,40,41,42,43,44,45,46}.  The data differ by a factor of two around the maximum. A recommended data set was also proposed earlier in the IAEA TLA library \cite{47}. In the presently investigated energy range experimental data are available from five groups (Albuoy, Michel 1983, Michel 1979, Hontzeas and Zhao)\cite{29, 30, 37, 48, 49}. Our data are in good agreement with the earlier results of Michel and Zhao (Fig. 1). Both the TENDL and the EMPIRE predictions describe the experimental data acceptable well.

\begin{figure}
\includegraphics[scale=0.3]{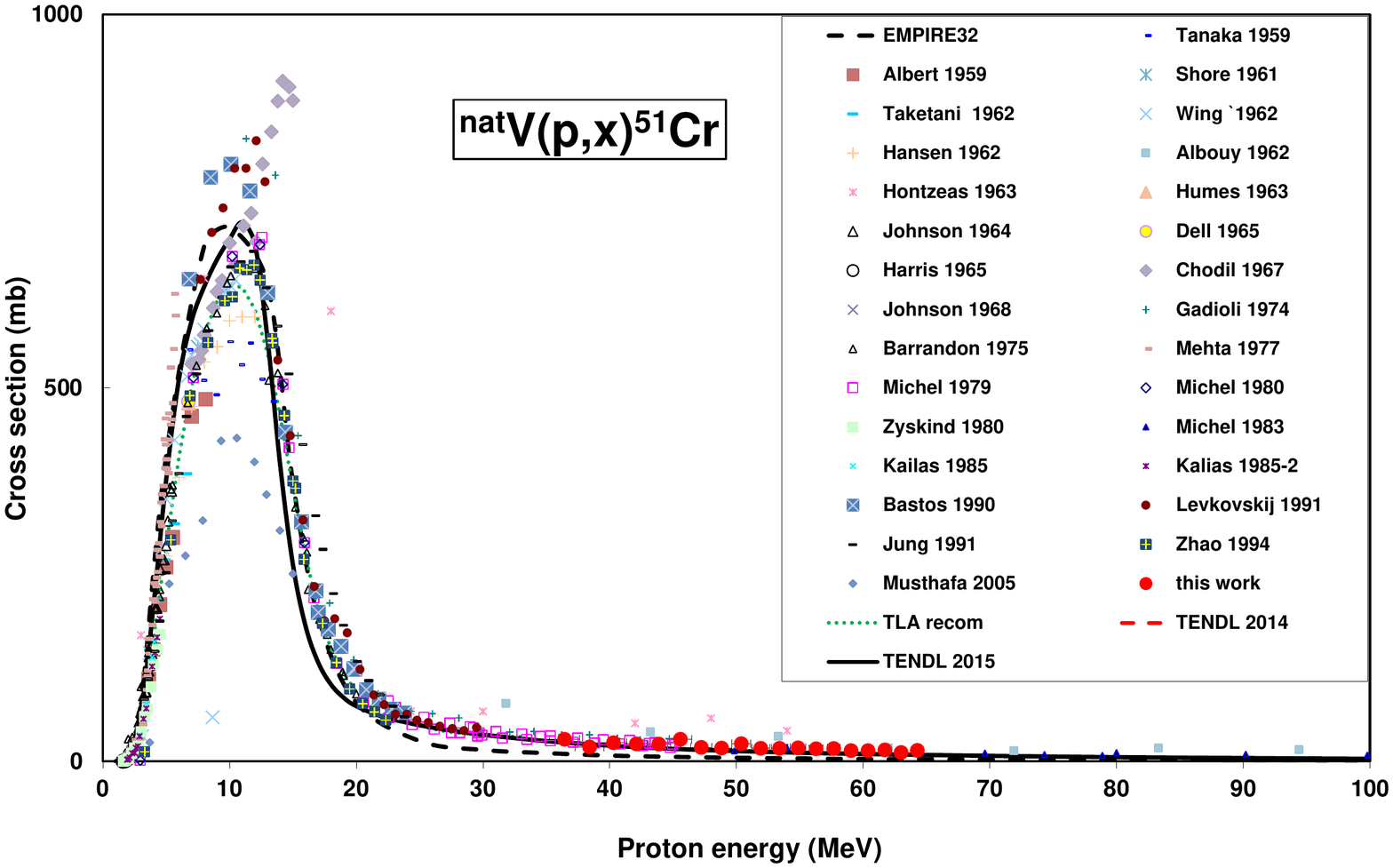}
\caption{Experimental excitation function for the $^{nat}$V(p,x)$^{51}$Cr reaction and comparison with literature values and theoretical code calculations}
\end{figure}

\subsection{Production of  $^{48}$Cr}
\label{4.2}

For production cross sections of $^{48}$Cr (T$_{1/2}$ = 21.56 h) the earlier experimental data of (Heininger, Hontzeas and Michel 1979)\cite{30, 37, 50} show large discrepancies (Fig. 2). Our data support the results of the Hannover group (Michel et al.)\cite{48}. The TENDL prediction shows only a moderate agreement (lower maximum, energy shift). A gradual change (increase) of the subsequent TENDL versions can be observed. The EMPIRE 3.2 results describe both the maximum and the magnitude of the excitation function well.

\begin{figure}
\includegraphics[scale=0.3]{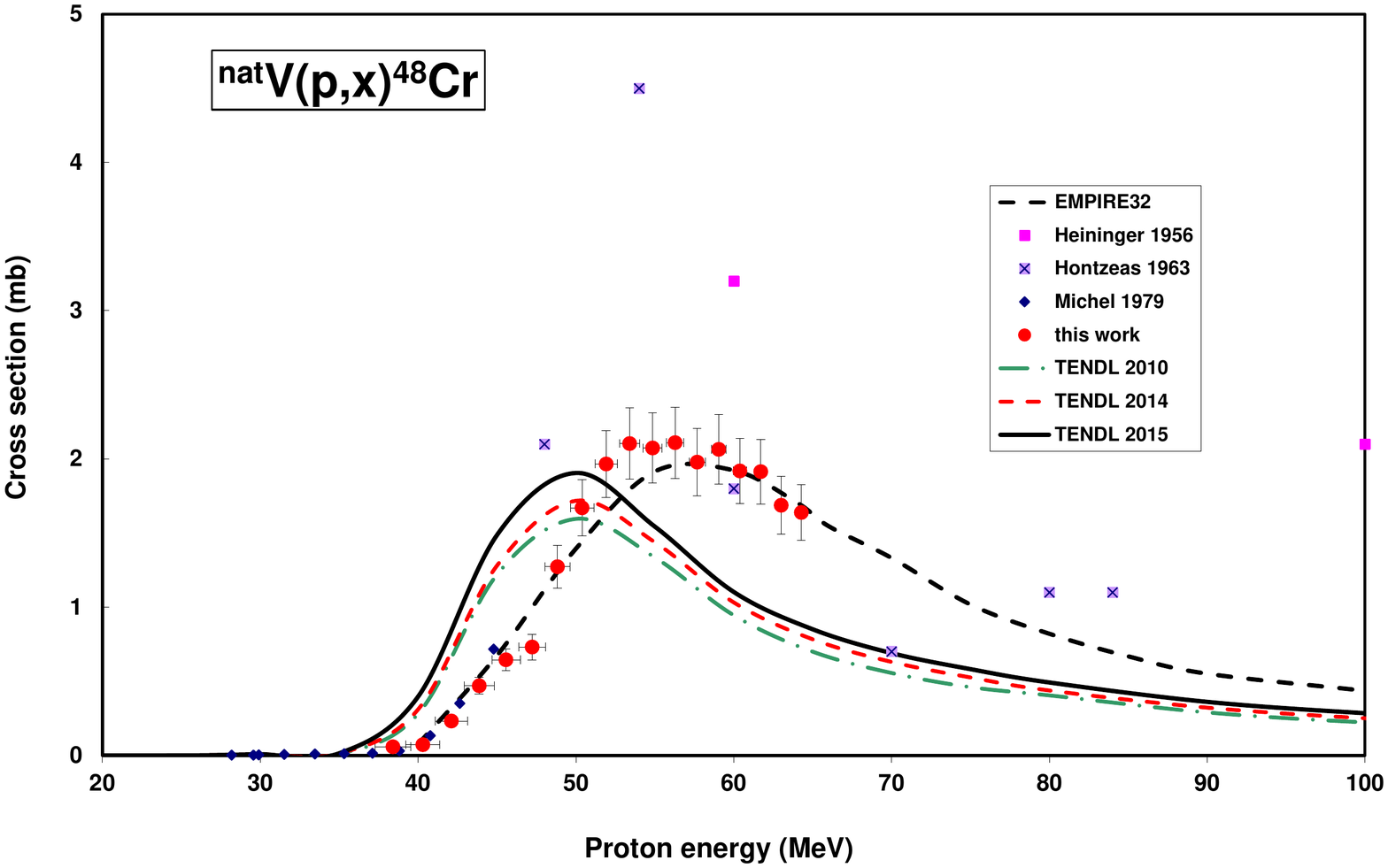}
\caption{Experimental excitation function for the $^{nat}$V(p,x)$^{48}$Cr reaction and comparison with literature values and theoretical code calculations}
\end{figure}

\subsection{Production of $^{48}$V}
\label{4.3}

The measured activation cross sections for $^{48}$V (T$_{1/2}$ = 15.9735 d) formation are cumulative. They include the direct production by (p,p3n) reaction with a threshold around 32 MeV and indirect production through the decay of the $^{48}$Cr (21.56 h) parent isotope.  The agreement with the earlier experimental data of Heininger, Hontzeas and Michel \cite{30, 37, 48, 50} is good and the TENDL data are slightly shifted with an overestimation of the maximum (Fig. 3). EMPIRE 3.2 gives better estimation for the maximum, but underestimates the experimental values above 50 MeV.

\begin{figure}
\includegraphics[scale=0.3]{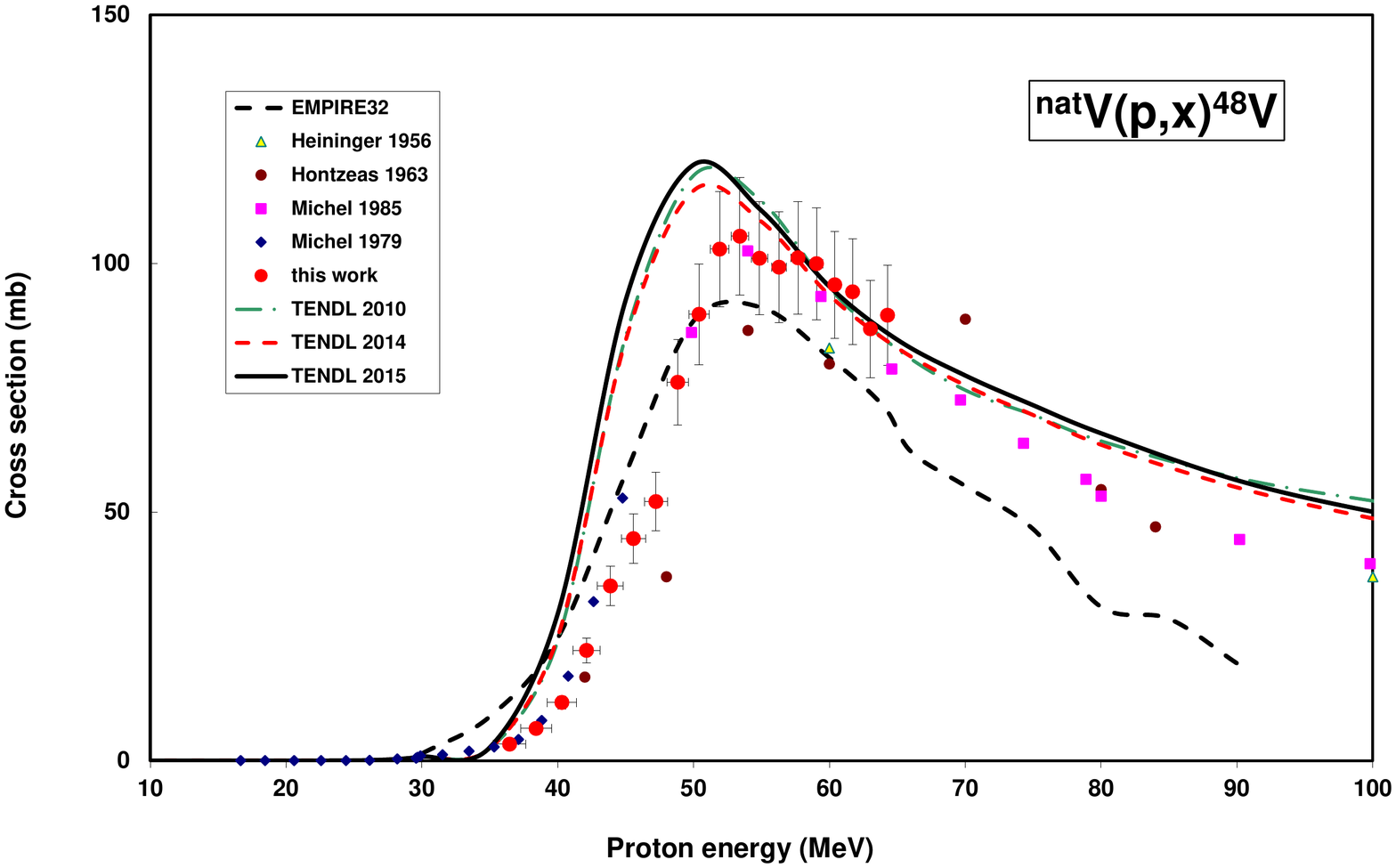}
\caption{Experimental excitation function for the $^{nat}$V(p,x)$^{48}$V reaction and comparison with literature values and theoretical code calculations}
\end{figure}

\subsection{Production of $^{48}$Sc}
\label{4.4}

The $^{48}$Sc (T$_{1/2}$ = 43.67 h) is produced directly, via (p,2pxn) reactions. The new experimental data are in good agreement with the earlier experimental data of the Hannover group (Michel)\cite{37, 48}. The data of Hontzeas \cite{30} however are significantly lower. The TENDL libraries overestimate the maximum (Fig. 4). EMPIRE gives strong overestimation.

\begin{figure}
\includegraphics[scale=0.3]{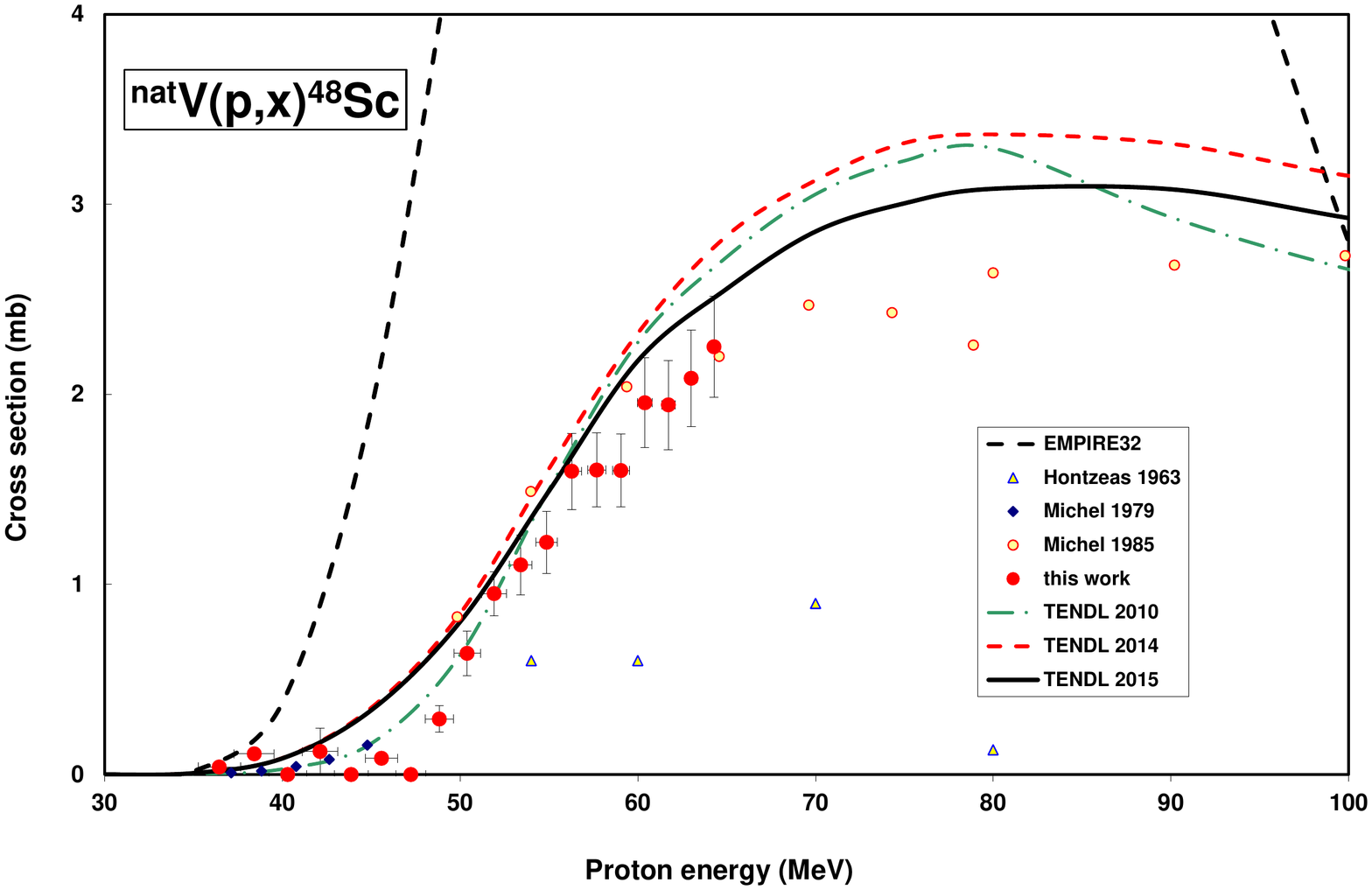}
\caption{Experimental excitation function for the $^{nat}$V(p,x)$^{48}$Sc reaction and comparison with literature values and theoretical code calculations}
\end{figure}

\subsection{Production of $^{47}$Sc}
\label{4.5}
The direct production cross sections of $^{47}$Sc(T$_{1/2}$ = 3.3492 d) are shown in Fig 5. The maximum around 38 MeV corresponds to the contribution of the $^{51}$V(p,$\alpha$p) reaction with energy threshold of 10.3 MeV, while the rise above 50 MeV is due to the emission of individual nucleons. A good agreement was found in the overlapping energy region with the experimental data of the Hannover group (Levkovskij, Michel 1979 and Michel 1985)\cite{37, 48}. The data of Levkovskij were normalized because of a known error in the original data-set \cite{51}. The data of Heininger and Hontzeas \cite{30, 50} differ significantly. Although the TENDL predicts well the shape of the excitation function but the values are underestimated below 50 MeV. EMPIRE 3.2 follows the shape of the experimental values but strongly overestimates them.

\begin{figure}
\includegraphics[scale=0.3]{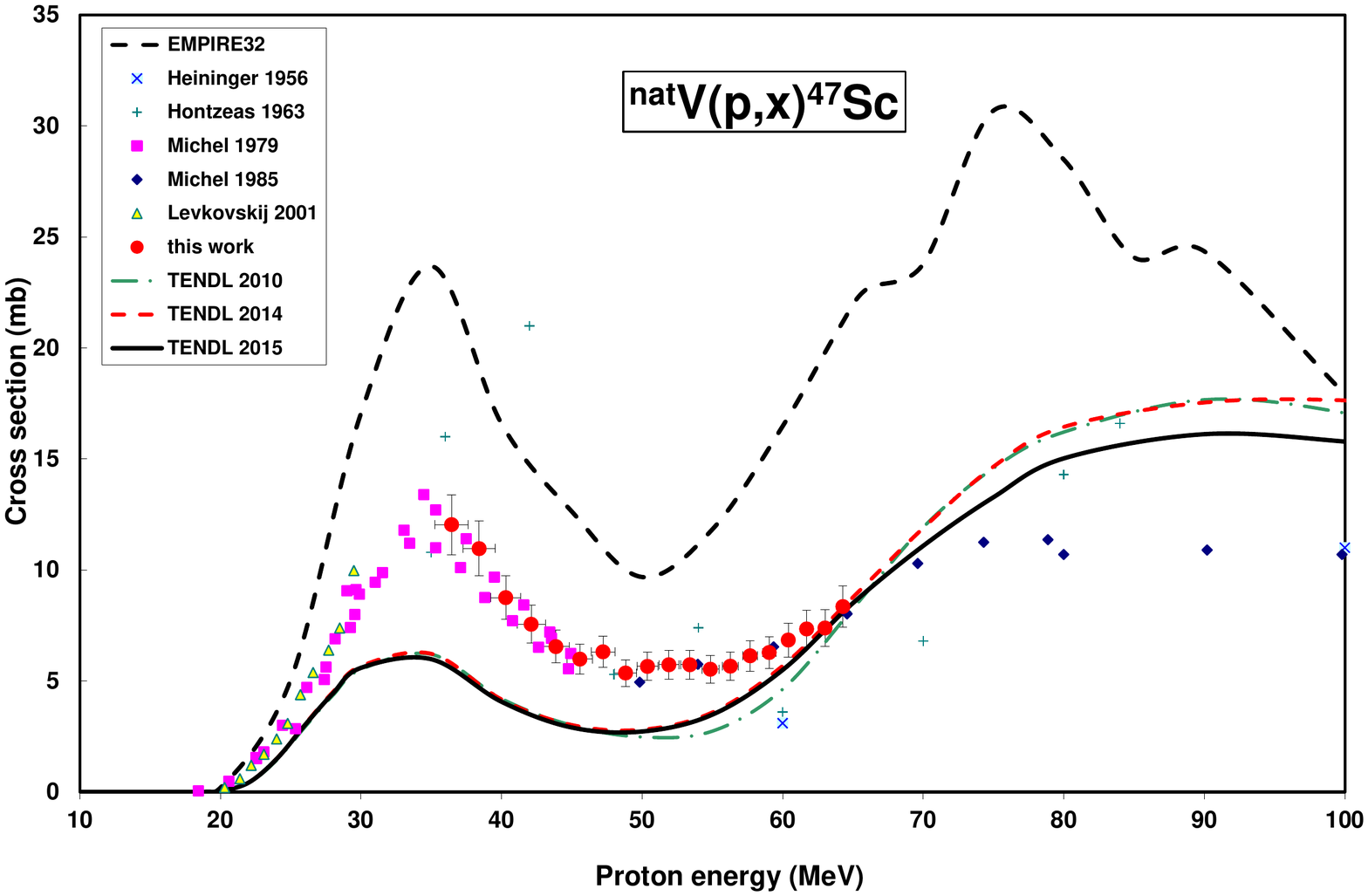}
\caption{Experimental excitation function for the $^{nat}$V(p,x) $^{47}$Screaction and comparison with literature values and theoretical code calculations}
\end{figure}

\subsection{Production of $^{46}$Sc}
\label{4.6}
The cross-sections of $^{46}$Sc(T$_{1/2}$ = 83.79 d) are cumulative as they include  direct production and total decay of the 19.7 s  half- life isomeric state. Up to 60 MeV the main contribution is from the $^{51}$V(p,$\alpha$pn) reaction. The comparison with the earlier experimental data shows good agreement with the data-sets of  Michel and Zaitseva \cite{37, 48, 52}, and more or less with Heininger \cite{50}(Fig. 6).  There are large disagreements with the data of  Albert and Hontzeas \cite{24, 30}. The TENDL predictions significantly underestimate the experimental data up to 80 MeV. The maximum position predicted by the EMPIRE 3.2 is better, but its overestimation is large.			

\begin{figure}
\includegraphics[scale=0.3]{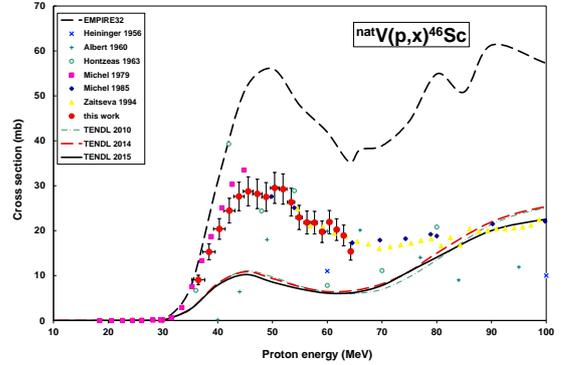}
\caption{Experimental excitation function for the $^{nat}$V(p,x)$^{46}$Sc  reaction and comparison with literature values and theoretical code calculations}
\end{figure}

\subsection{Production of $^{44m}$Sc and $^{44g}$Sc}
\label{4.7}
The radionuclide $^{44}$Sc has two long-lived isomeric states. The metastable $^{44m}$Sc(T$_{1/2}$ = 58.61 h) is produced directly and decays with 98.8 \% IT to the ground state and the rest 1.2 \% is EC + $\beta^+$. The ground state $^{44g}$Sc (T$_{1/2}$ = 3.97 h) is produced directly, through decay of its metastable state (58.61 h) and through the decay of its long half-life (T$_{1/2}$ = 44.3 a) $^{44}$Ti parent.
Direct production cross-sections for the ground state were deduced after correction of the contribution of $^{44m}$Sc decay. The possible contribution of the $^{44}$Ti was neglected because of the irradiation conditions and its very long half-life that would not generate any significant signal in our experimental conditions.  For both states reactions with emissions of clusters are playing the dominant role in the energy domain studied. The measured and calculated cross sections are shown in Fig. 7 and Fig. 8. 
In case of $^{44m}$Sc our new data agree well with the earlier experimental data of Michel \cite{48}. The data  of Hontzeas and Heininger \cite{30, 50} differ significantly. The TENDL predictions show a significant underestimation (see Fig. 7). EMPIRE 3.2 gives much better approximation both for the values and for the shape.
For production of $^{44g}$Sc only one earlier experimental cross-section data set was found  \cite{30}, which is in good agreement with our results. The predictions of TENDL are representing well the experimental values (Fig. 8). EMPIRE 3.2 strongly overestimates the experimental values.

\begin{figure}
\includegraphics[scale=0.3]{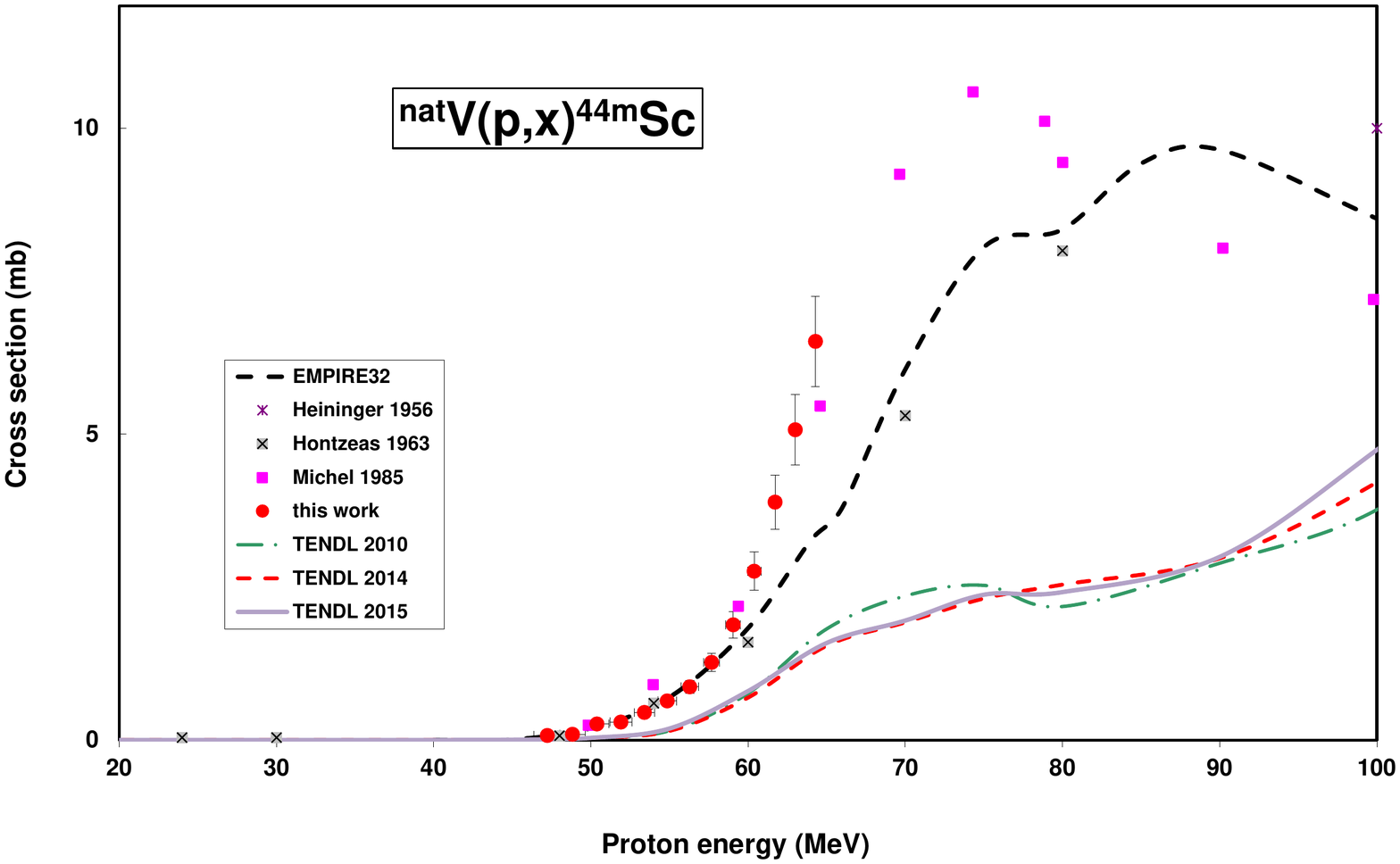}
\caption{Experimental excitation function for the $^{nat}$V(p,x)$^{44m}$Sc  reaction and comparison with literature values and theoretical code calculations}
\end{figure}

\begin{figure}
\includegraphics[scale=0.3]{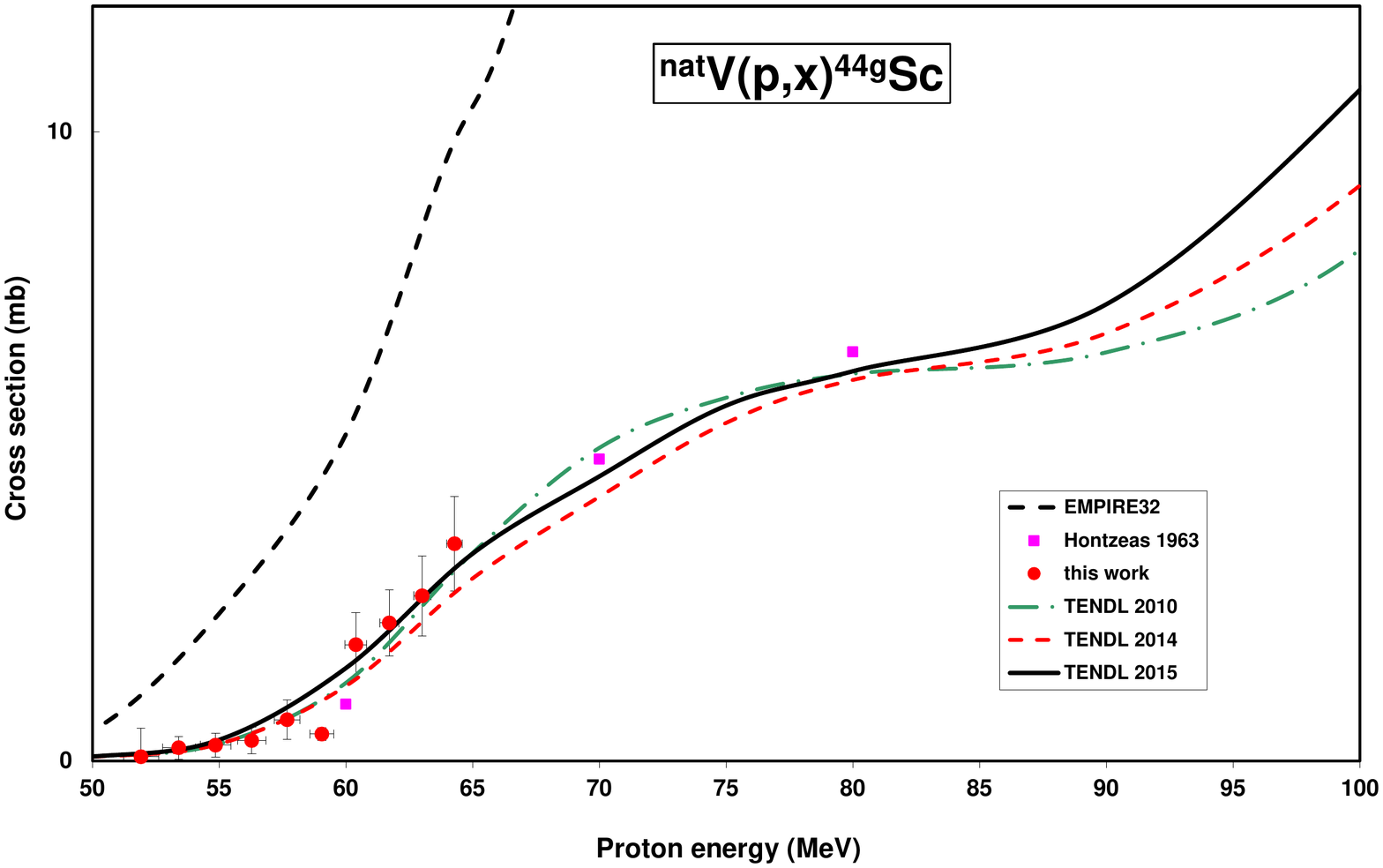}
\caption{Experimental excitation function for the $^{nat}$V(p,x)$^{44g}$Sc  reaction and comparison with literature values and theoretical code calculations}
\end{figure}

\subsection{Production of $^{43}$Sc}
\label{4.8}
The cumulative cross-sections of  $^{43}$Sc (T$_{1/2}$ = 3.891 h), by direct production and by the possible decay of the short-lived $^{43}$Ti (T$_{1/2}$ = 509 ms, formed through the $^{51}$V(p,$\alpha$5n) reaction with a threshold of  59.53 MeV), are shown in Fig. 9 with earlier data of Hontzeas \cite{30} and with the TENDL calculations. Our data are significantly higher than those of Hontzeas. The TENDL data are even lower and the effective threshold is also shifted (Fig. 9). EMPIRE 3.2 gives similar underestimation, but closer to the experimental values.

\begin{figure}
\includegraphics[scale=0.3]{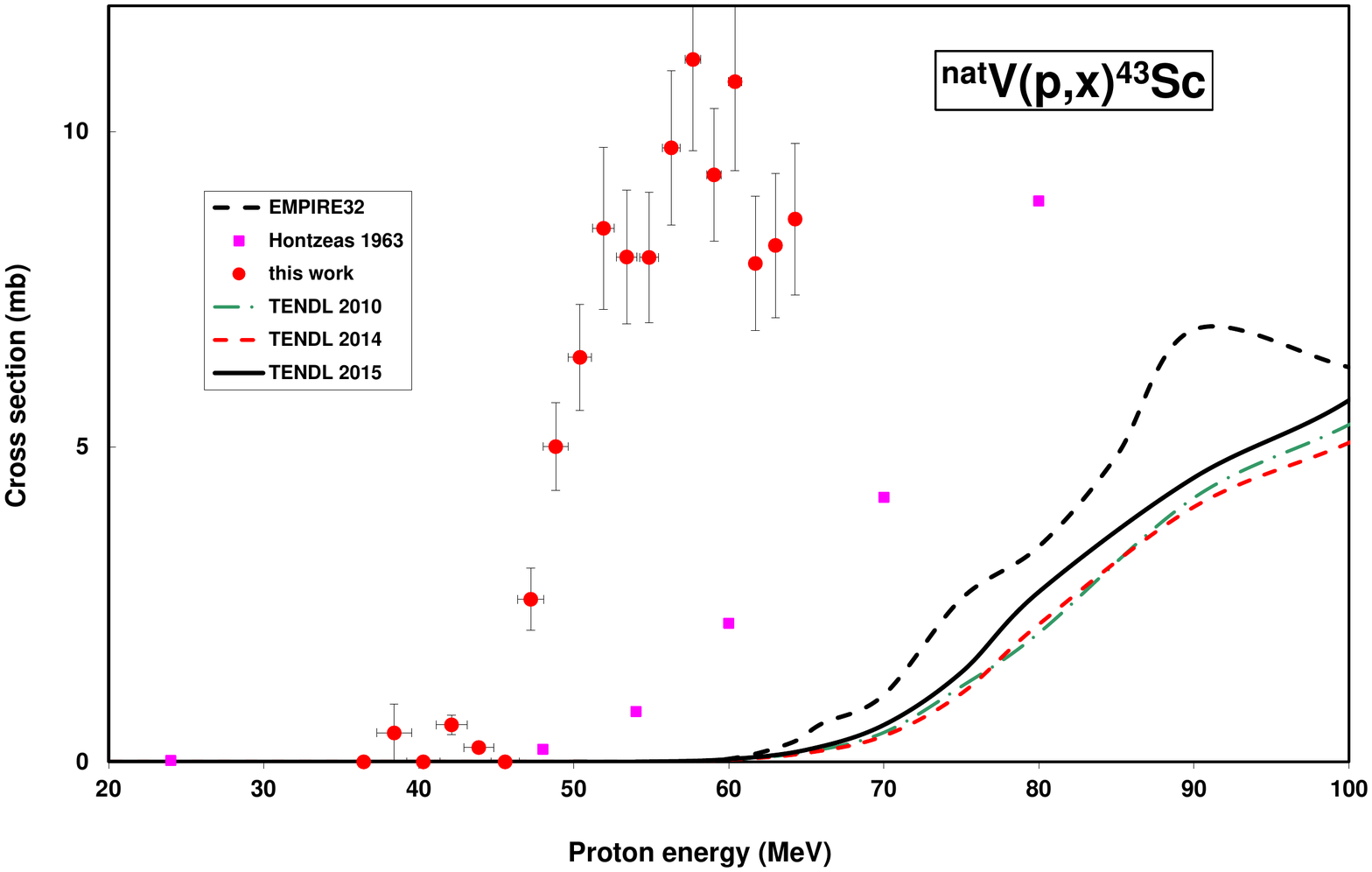}
\caption{Experimental excitation function for the $^{nat}$V(p,x)$^{43}$Sc reaction and comparison with literature values and theoretical code calculations}
\end{figure}

\subsection{Production of $^{43}$K}
\label{4.9}
The direct production cross-sections of $^{43}$K (T$_{1/2}$ = 2.3 h), essentially formed through clustered emissions in our experimental conditions, are shown in Fig. 10. The agreement with the data of Michel \cite{48} is good. The data of Heininger and Hontzeas \cite{30, 50} are higher. The underestimation by the TENDL prediction is significant. (Fig. 10). EMPIRE 3.2 gives better result with overestimation up to 50 MeV, above this energy the overestimation is much larger.

\begin{figure}
\includegraphics[scale=0.3]{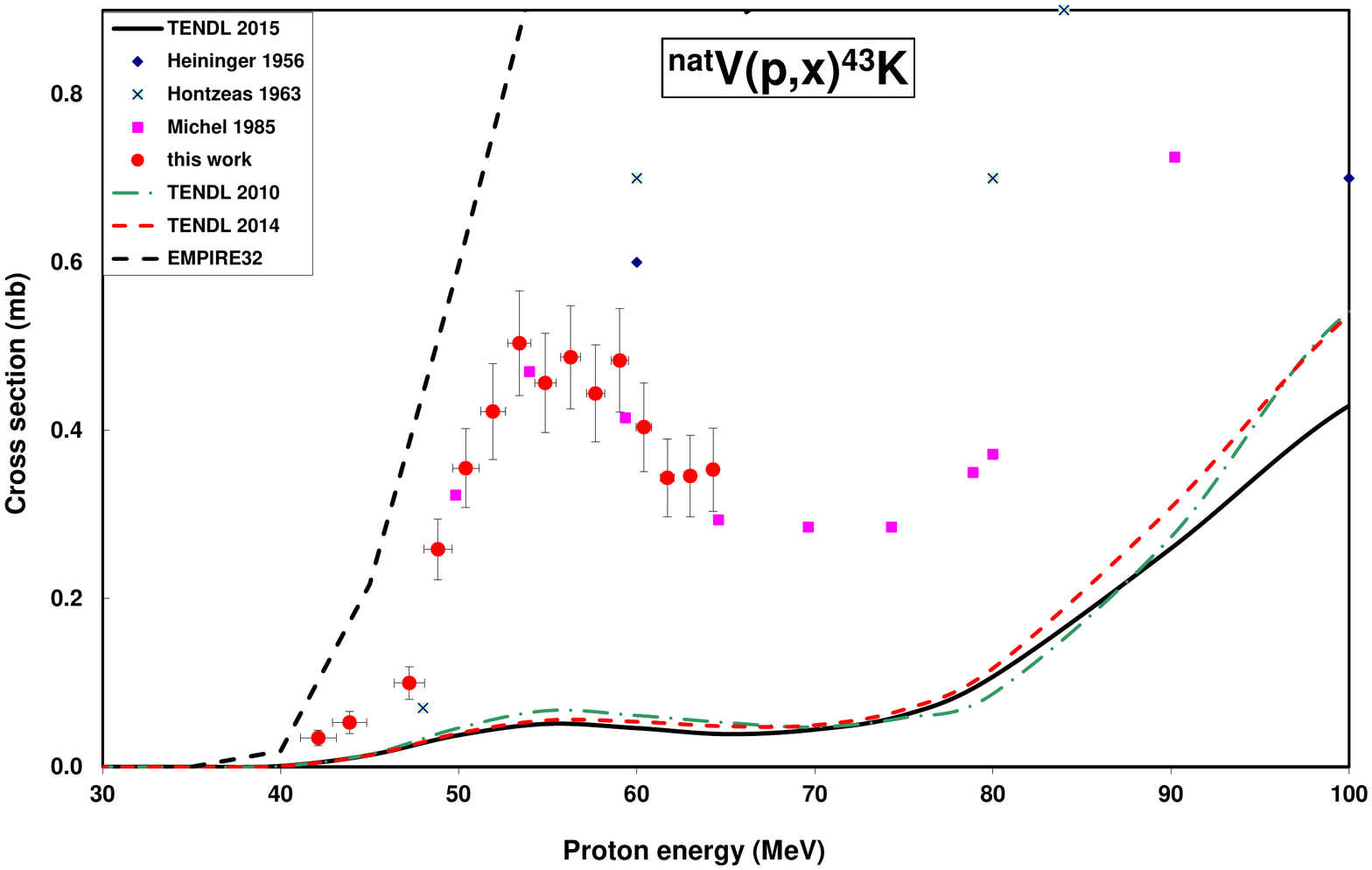}
\caption{Experimental excitation function for the $^{nat}$V(p,x)$^{43}$K reaction and comparison with literature values and theoretical code calculations}
\end{figure}

\subsection{Production of $^{42}$K}
\label{4.10}
The $^{42}$K (T$_{1/2}$ = 12.36 h) is produced directly through clustered emission. The comparison shows good agreement with the results of Michel 1985 \cite{48}. The data of Heininger and Hontzeas \cite{30, 50} are significantly lower in the measured energy range. The situation is the same for the TENDL predictions (Fig. 11) as in the previous case. For EMPIRE 3.2 the same behavior can be observed as in the previous case, acceptable agreement with slight overestimation up to 60 MeV, and strong overestimation above.

\begin{figure}
\includegraphics[scale=0.3]{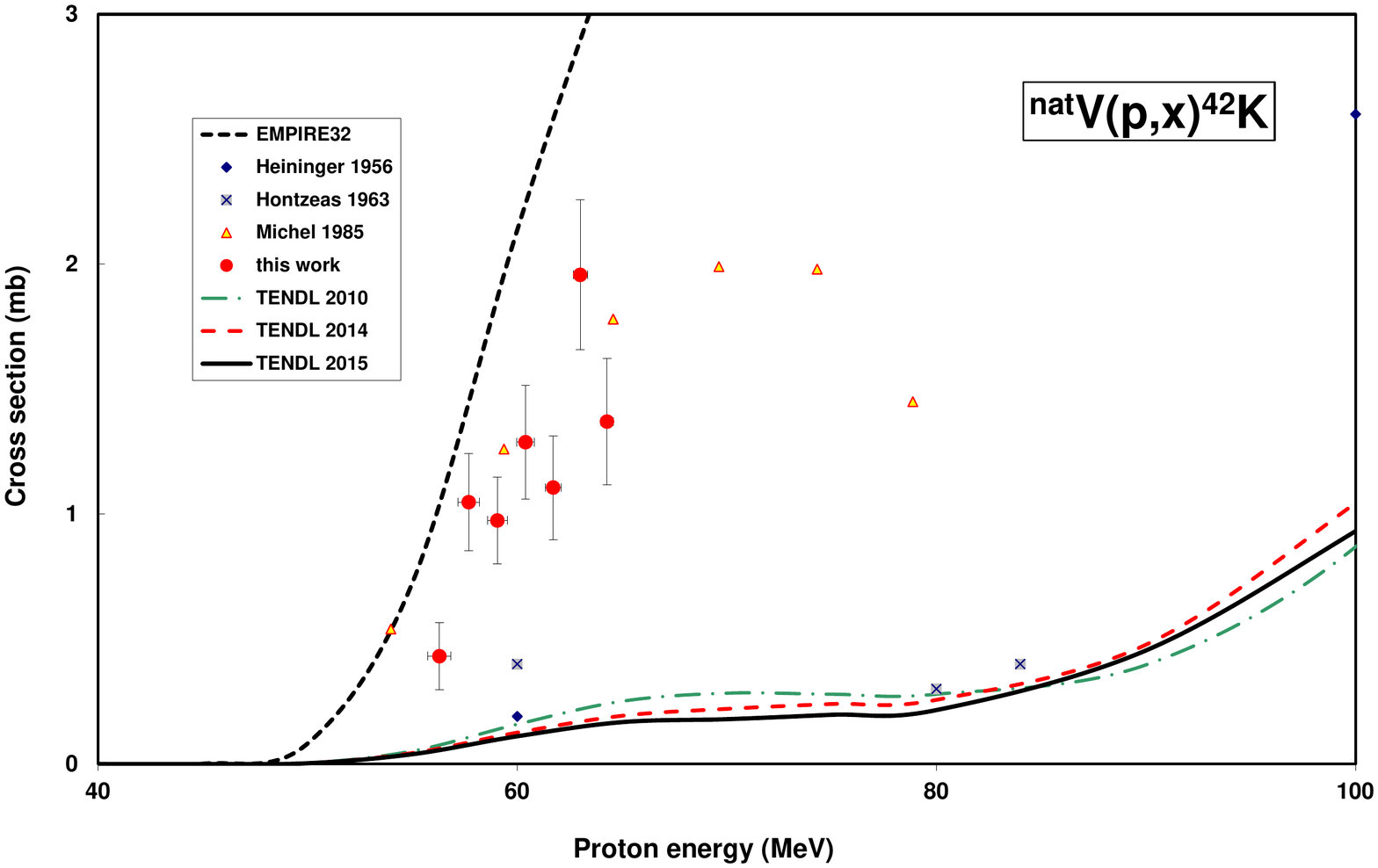}
\caption{Experimental excitation function for the $^{nat}$V(p,x)$^{42}$K reaction and comparison with literature values and theoretical code calculations}
\end{figure}

\begin{table*}[t]
\tiny
\caption{Measured production cross section of radioisotopes from the $^{nat}$V(p,x) reactions}
\centering
\begin{center}
\begin{tabular}{|p{0.1in}|p{0.1in}|p{0.1in}|p{0.1in}|p{0.1in}|p{0.1in}|p{0.1in}|p{0.1in}|p{0.1in}|p{0.1in}|p{0.1in}|p{0.1in}|p{0.1in}|p{0.1in}|p{0.1in}|p{0.1in}|p{0.1in}|p{0.1in}|p{0.1in}|p{0.1in}|p{0.1in}|p{0.1in}|p{0.1in}|p{0.1in}|} \hline 
\multicolumn{2}{|c|}{\textbf{Energy}} & \multicolumn{2}{|c|}{\textbf{${}^{51}$Cr}} & \multicolumn{2}{|c|}{\textbf{${}^{48}$Cr}} & \multicolumn{2}{|c|}{\textbf{$^{48}$V}} & \multicolumn{2}{|c|}{\textbf{$^{48}$Sc}} & \multicolumn{2}{|c|}{\textbf{$^{47}$Sc}} & \multicolumn{2}{|c|}{\textbf{$^{46}$Sc}} & \multicolumn{2}{|c|}{\textbf{$^{44m}$Sc}} & \multicolumn{2}{|c|}{\textbf{$^{44g}$Sc}} & \multicolumn{2}{|c|}{\textbf{$^{43}$Sc}} & \multicolumn{2}{|c|}{\textbf{$^{43}$K}} & \multicolumn{2}{|c|}{\textbf{$^{42}$K}} \\ \hline 
\multicolumn{2}{|c|}{\textbf{MeV}} & \multicolumn{22}{|c|}{\textbf{Cross section (mb)}} \\ \hline 
\textbf{64.28} & \textbf{0.30} & 14.88 & 2.19 & 1.64 & 0.19 & 89.59 & 10.09 & 2.25 & 0.26 & 8.35 & 0.94 & 15.37 & 1.94 & 6.52 & 0.74 & 3.45 & 0.76 & 8.61 & 1.21 & 0.35 & 0.05 & 1.37 & 0.25 \\ \hline 
\textbf{63.01} & \textbf{0.34} & 12.00 & 1.93 & 1.69 & 0.19 & 86.83 & 9.78 & 2.08 & 0.25 & 7.39 & 0.83 & 18.92 & 2.31 & 5.07 & 0.57 & 2.62 & 0.63 & 8.20 & 1.14 & 0.35 & 0.05 & 1.96 & 0.30 \\ \hline 
\textbf{61.71} & \textbf{0.38} & 15.40 & 2.12 & 1.91 & 0.22 & 94.32 & 10.61 & 1.94 & 0.23 & 7.35 & 0.83 & 20.24 & 2.41 & 3.89 & 0.44 & 2.19 & 0.52 & 7.91 & 1.06 & 0.34 & 0.05 & 1.11 & 0.21 \\ \hline 
\textbf{60.39} & \textbf{0.42} & 14.41 & 2.00 & 1.92 & 0.22 & 95.70 & 10.76 & 1.96 & 0.24 & 6.84 & 0.77 & 21.91 & 2.59 & 2.76 & 0.31 & 1.85 & 0.51 & 10.80 & 1.41 & 0.40 & 0.05 & 1.29 & 0.23 \\ \hline 
\textbf{59.05} & \textbf{0.47} & 14.41 & 2.09 & 2.07 & 0.24 & 99.94 & 11.24 & 1.60 & 0.19 & 6.28 & 0.71 & 19.75 & 2.38 & 1.88 & 0.22 & 0.43 & 0.10 & 9.32 & 1.05 & 0.48 & 0.06 & 0.97 & 0.17 \\ \hline 
\textbf{57.68} & \textbf{0.51} & 17.14 & 2.14 & 1.98 & 0.23 & 101.11 & 11.36 & 1.60 & 0.19 & 6.13 & 0.69 & 21.77 & 2.52 & 1.27 & 0.15 & 0.65 & 0.31 & 11.15 & 1.45 & 0.44 & 0.06 & 1.05 & 0.20 \\ \hline 
\textbf{56.29} & \textbf{0.56} & 16.83 & 2.20 & 2.11 & 0.24 & 99.26 & 11.16 & 1.60 & 0.20 & 5.67 & 0.64 & 21.81 & 2.56 & 0.87 & 0.10 & 0.32 & 0.21 & 9.75 & 1.23 & 0.49 & 0.06 & 0.43 & 0.13 \\ \hline 
\textbf{54.86} & \textbf{0.60} & 18.14 & 2.31 & 2.07 & 0.24 & 101.02 & 11.36 & 1.22 & 0.16 & 5.53 & 0.62 & 22.94 & 2.68 & 0.64 & 0.08 & 0.25 & 0.19 & 8.00 & 1.03 & 0.46 & 0.06 & ~ &  \\ \hline 
\textbf{53.41} & \textbf{0.65} & 17.61 & 2.40 & 2.10 & 0.24 & 105.46 & 11.86 & 1.10 & 0.16 & 5.73 & 0.65 & 26.34 & 3.09 & 0.45 & 0.07 & 0.21 & 0.18 & 8.01 & 1.06 & 0.50 & 0.06 & ~ &  \\ \hline 
\textbf{51.92} & \textbf{0.70} & 17.57 & 2.26 & 1.97 & 0.23 & 102.89 & 11.57 & 0.95 & 0.12 & 5.73 & 0.64 & 29.26 & 3.38 & 0.29 & 0.04 & 0.07 & 0.45 & 8.47 & 1.29 & 0.42 & 0.06 & ~ &  \\ \hline 
\textbf{50.39} & \textbf{0.74} & 23.67 & 3.03 & 1.67 & 0.19 & 89.76 & 10.10 & 0.64 & 0.12 & 5.66 & 0.64 & 29.50 & 3.45 & 0.26 & 0.06 & ~ &  & 6.42 & 0.84 & 0.36 & 0.05 & ~ &  \\ \hline 
\textbf{48.83} & \textbf{0.79} & 17.82 & 2.28 & 1.27 & 0.15 & 76.12 & 8.57 & 0.29 & 0.07 & 5.36 & 0.60 & 27.51 & 3.20 & 0.09 & 0.03 & ~ &  & 5.00 & 0.69 & 0.26 & 0.04 & ~ &  \\ \hline 
\textbf{47.22} & \textbf{0.85} & 18.91 & 2.33 & 0.73 & 0.09 & 52.09 & 5.87 &  &  & 6.32 & 0.71 & 28.18 & 3.26 & 0.08 & 0.02 & ~ &  & 2.58 & 0.49 & 0.10 & 0.02 & ~ &  \\ \hline 
\textbf{45.57} & \textbf{0.90} & 29.48 & 3.34 & 0.65 & 0.07 & 44.64 & 5.01 & 0.09 & 0.01 & 5.99 & 0.67 & 28.75 & 3.24 & ~ &  & ~ &  & ~ &  & ~ &  & ~ &  \\ \hline 
\textbf{43.87} & \textbf{0.95} & 23.15 & 2.72 & 0.47 & 0.06 & 35.15 & 3.96 &  &  & 6.55 & 0.74 & 27.61 & 3.16 & ~ &  & ~ &  & 0.23 & 0.06 & 0.05 & 0.01 & ~ &  \\ \hline 
\textbf{42.11} & \textbf{1.01} & 23.31 & 2.63 & 0.23 & 0.03 & 22.20 & 2.49 & 0.12 & 0.12 & 7.56 & 0.85 & 24.45 & 2.75 & ~ &  & ~ &  & 0.58 & 0.16 & 0.03 & 0.01 & ~ &  \\ \hline 
\textbf{40.30} & \textbf{1.07} & 24.97 & 2.84 & 0.07 & 0.01 & 11.70 & 1.32 &  &  & 8.75 & 0.98 & 20.38 & 2.31 & ~ &  & ~ &  & ~ &  & ~ &  & ~ &  \\ \hline 
\textbf{38.41} & \textbf{1.13} & 19.35 & 2.24 & 0.06 & 0.03 & 6.46 & 0.78 & 0.11 & 0.03 & 10.97 & 1.23 & 29.73 & 3.37 & ~ &  & ~ &  & 0.46 & 0.46 & ~ &  & ~ &  \\ \hline 
\textbf{36.44} & \textbf{1.19} & 29.51 & 3.34 & 0.01 & 0.00 & 3.31 & 0.38 & 0.04 & 0.01 & 12.04 & 1.35 & 9.04 & 1.04 & ~ &  & ~ &  & ~ &  & ~ &  & ~ &  \\ \hline 
\end{tabular}

\end{center}
\end{table*}

\section{Thick target yields}
\label{5}
From excitation functions obtained by a spline fit to our experimental data, integral thick target yields \cite{53} were calculated and are shown in Figs. 12 and 13 as a function of the incident  proton energy. The deduced integral yields are compared with the experimental thick target yields from the literature measured by Dmitriev, Acerbi, Nickles, Abe, Bastos, L. Kanza-Kanza, etc.\cite{5,6,7,8, 41, 54, 55}, where overlapping energy regions made it possible. 
Our calculated $^{51}$Cr yield is in good agreement with the yields of Dmitriev and Bastos \cite{41, 55}, much higher than the yield of Abe \cite{7}, slightly higher than the yield of Nickles \cite{8} and slightly lower than the yield curve of Acerbi \cite{6}. Our $^{48}$V yield is in excellent agreement with the yield curve of Acerbi \cite{6}, but Dmitriev and Molin \cite{55} reported much higher yield at 22 MeV than our value. For the other isotopes in Fig. 12 ($^{48}$Cr and $^{48}$Sc) there were no literature yield data found. 
In Fig. 13 our calculated yields for $^{46,44m,44g,43}$Sc and $^{43,42}$K are presented. Previous data in the literature were only found for $^{46}$Sc (Acerbi). Their data are in excellent agreement with our new results.

\begin{figure}
\includegraphics[scale=0.3]{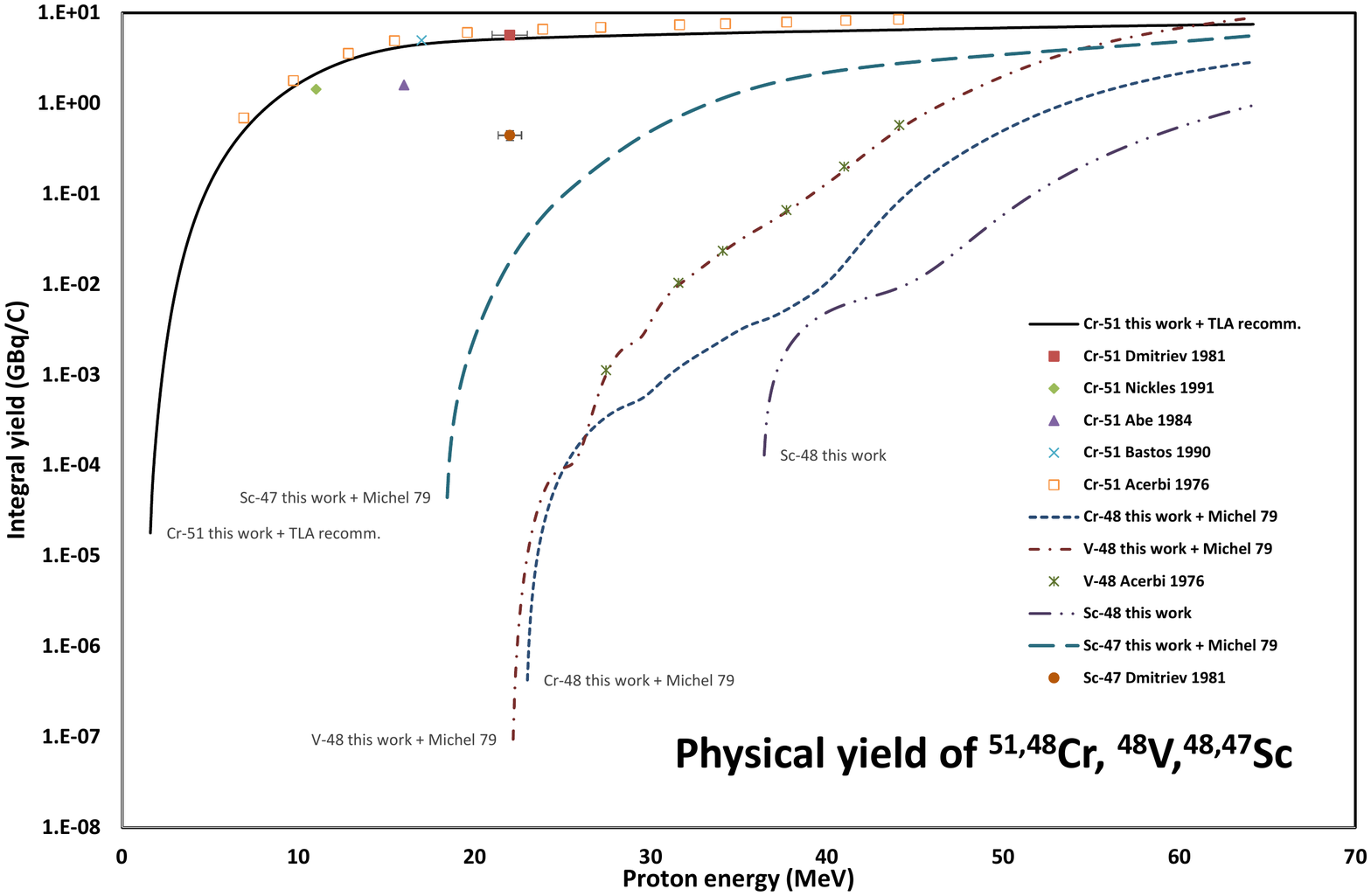}
\caption{Integral yields of the $^{51,48}$Cr, $^{48}$V and $^{48,47}$Sc calculated from the fitted experimental excitation functions and compared with the literature data}
\end{figure}

\begin{figure}
\includegraphics[scale=0.3]{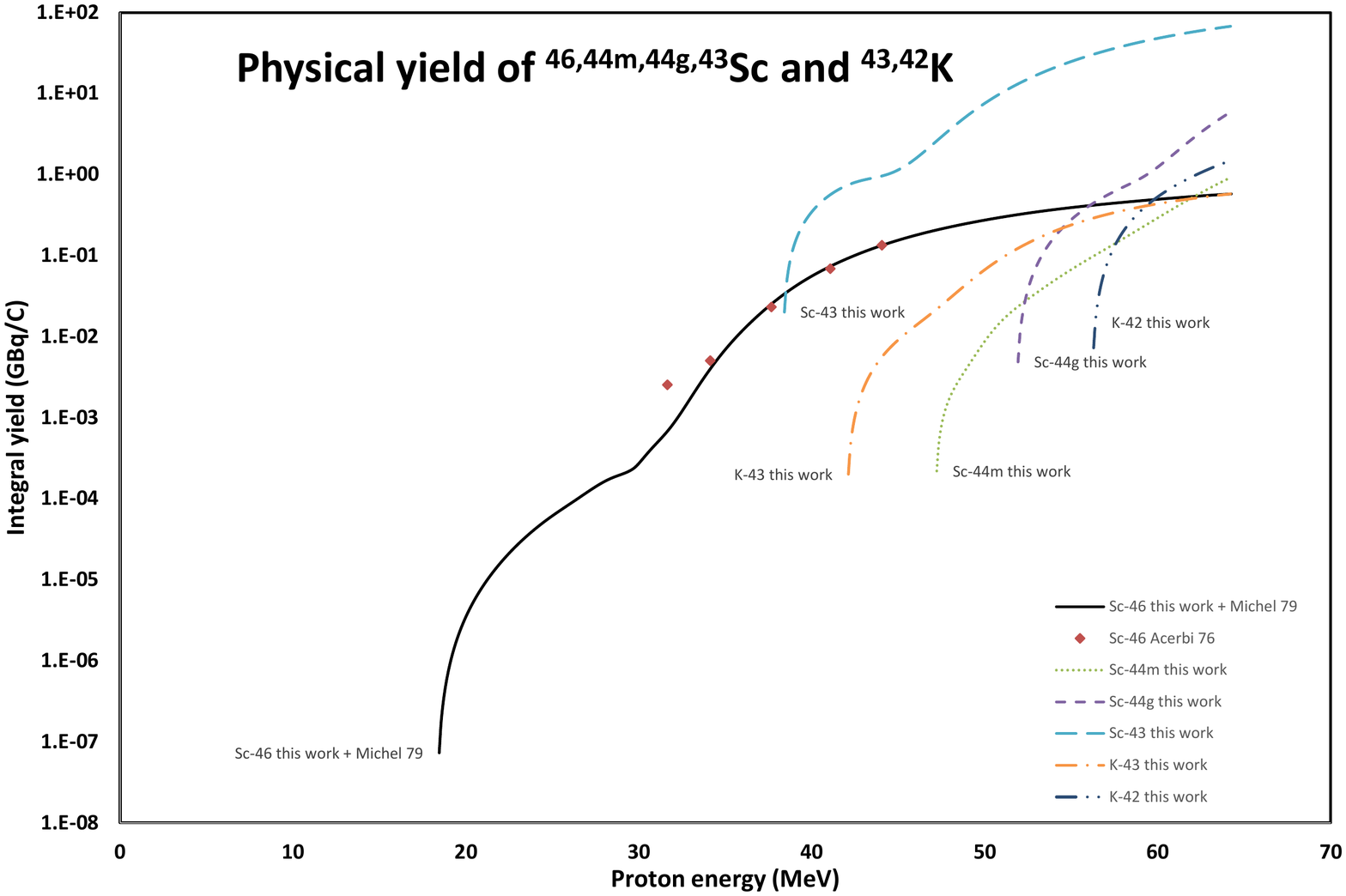}
\caption{Integral yields of the $^{46,44m,44g,43}$Sc and $^{43,42}$K calculated from the fitted experimental excitation functions and compared with the literature data}
\end{figure}

\section{Applications}
\label{6}

\subsection{Thin layer activation (TLA)}
\label{6.1}

Among the investigated reaction products, the production and decay parameters of $^{51}$Cr (T$_{1/2}$ = 27.7010 d, E$_{\gamma}$ = 320.0824 keV (9.91 \%)) are the most suitable for satisfying the usual requirements (medium half-lives, intensive gamma-lines, high production rate) \cite{56, 57} for application with TLA. Recommended cross sections and specific activity as a function of depth up to 20 MeV can also be found in the IAEA TLA \cite{47} database for the $^{nat}$V(p,x)$^{51}$Cr reaction. The recommended data are shown in Fig. 1 together with the experimental data \cite{47}. The fitted curve follows reasonably well the more reliable selected data. The corresponding wear curve (specific activity versus depth) is already presented in our previous work on deuteron induced reactions on vanadium \cite{58}. Fig. 14 demonstrates the capability of the method in this higher energy region, where other radioisotopes become suitable for wear measurement. As an example, wear curves for $^{48}$V (shorter, but proper half-life, intense $\gamma$-energies, cross section in this energy range is lower than that of the $^{51}$Cr around 11 MeV, but it is still applicable) are presented in Fig. 14. The used matrix is pure vanadium, because its interpretation for an alloy is straightforward. Both wear curves were constructed assuming 1 hour 1 $\mu$A irradiation with 50 MeV proton beam (recommended to achieve linear activity distribution near to the surface), under irradiation angles of 15$^o$ and 90$^o$ (perpendicular) respectively. In the first case 140 $\mu$m linear (within 1 \%) distribution and about 800 $\mu$m total penetration depth can be reached, while in the case of perpendicular irradiation these values are 540 $\mu$m and more than 2 mm respectively. Depending on the required task one can choose the best irradiation angle in order to cover the selected wear-depth in question with constant \cite{58} or linear activity distribution. If the tracing isotope does not have a local maximum in the selected energy range or lower penetration depth (lower irradiation energy) is required, the depth distribution should be set as “linear”, which requires the knowledge of the depth distribution function.
For long term studies and higher wear rates, besides the $^{48}$V, the $^{46}$Sc is also a proper choice because of its 83.79 d half-life and intense gamma-energies. Unfortunately, its production also requires a high energy accelerator, which is not everywhere available.

\begin{figure}
\includegraphics[scale=0.3]{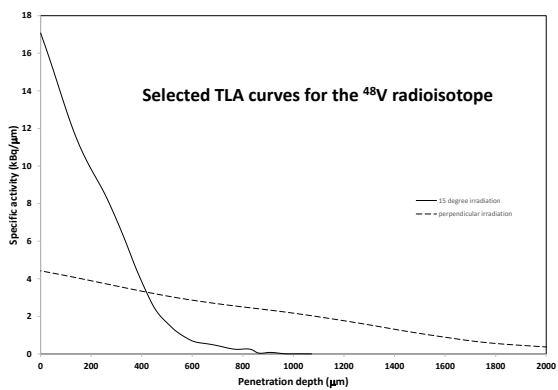}
\caption{TLA curves for the $^{48}$V radioisotope in natural vanadium matrix: irradiation time: 1 h, beam current: 1 $\mu$A, bombarding proton energy: 50 MeV, cooling time: 10 days}
\end{figure}

\section{Summary and conclusion}
\label{7}
Eleven excitation functions for proton induced nuclear reactions on natural vanadium were measured up to 65 MeV. Our new experimental data were critically compared with the values measured earlier by different authors. The experimental data sets show excellent agreement with the experimental data of the Michel group \cite{37, 38, 48} and with the Levkovskij and Zaitseva \cite{42, 52} data in the overlapping energy region. The comparison with the theoretical model code calculations as available in the on-line library TENDL (2014, 2015, except 2010) shows agreement for radio-products of chromium formed through (p,xn) reactions. In case of vanadium radionuclides, especially for scandium and potassium activation products, the underestimation is significant, but an improvement can be observed against the earlier (2010) version. EMPIRE 3.2 (Malta) provides in some cases much better approximation but fails in cases, where the produced isotope is far from the target isotope (multiple path or complex particle emission play role). The provided physical yield curves help the user to select the suitable isotope and reaction for the required medical isotope production or industrial task. The applicability of some isotopes, especially the $^{48}$V for wear measurements by using thin layer activation is demonstrated.

\section{Acknowledgements}

This work was done in the frame of MTA-FWO (Vlaanderen) research projects. The authors acknowledge the support of research projects and of their respective institutions in providing the materials and the facilities for this work. 
 



\bibliographystyle{elsarticle-num}
\bibliography{Vp}

\begin{thebibliography}{10}
\expandafter\ifx\csname url\endcsname\relax
  \def\url#1{\texttt{#1}}\fi
\expandafter\ifx\csname urlprefix\endcsname\relax\def\urlprefix{URL }\fi
\expandafter\ifx\csname href\endcsname\relax
  \def\href#1#2{#2} \def\path#1{#1}\fi

\bibitem{1}
C.~H. Johnson, A.~Galonsky, J.~P. Ulrich, Proton strength functions from (p,n),
  Phys. Rev. 109 (1958) 1243.

\bibitem{2}
C.~H. Johnson, C.~C. Trail, A.~Galonsky, Thresholds for (p,n) reactions on 26
  intermediate- weight nuclei, Phys. Rev. B 136 (1964) 1719.

\bibitem{3}
J.~L. Zyskind, Competition effects in charged particle induced reactions, Diss.
  Abst. Int. B 40 (1979) 1237.

\bibitem{4}
A.~H.~M. Solieman, M.~Al-Abyad, F.~Ditr\'oi, Z.~A. Saleh, Experimental and
  theoretical study for the production of 51cr using p, d, 3he and 4he
  projectiles on v, ti and cr targets, Nuclear Instruments \& Methods in
  Physics Research Section B 366 (2016) 19--27.

\bibitem{5}
P.~P. Dmitriev, Systematics of nuclear reaction yields for thick target at 22
  mev proton energy, Vop. At. Nauki i Tekhn., Ser.Yadernye Konstanty 57 (1983)
  2.

\bibitem{6}
E.~Acerbi, C.~Birattari, M.~Castiglioni, F.~Resmini, Nuclear applied physics at
  milan cyclotron, Journal of Radioanalytical Chemistry 34~(1) (1976) 191--217.

\bibitem{7}
K.~Abe, A.~Iizuka, A.~Hasegawa, S.~Morozumi, Induced radioactivity of component
  materials by 16-mev protons and 30-mev alpha-particles, Journal of Nuclear
  Materials 123~(1-3) (1984) 972--976.

\bibitem{8}
R.~J. Nickles, A shotgun approach to the chart of the nuclides, Acta Radiol.
  Suppl. 376 (1991) 69--71.

\bibitem{9}
L.~F. Mausner, K.~L. Kolsky, Radionuclide development at bnl for nuclear
  medicine therapy, Applied Radiation and Isotopes 49 (1998) 285.

\bibitem{10}
S.~M. Qaim, F.~T\'ark\'anyi, R.~Capote, Nuclear data for the production of
  therapeutic radionuclides, Tech. rep., IAEA (2011).

\bibitem{11}
C.~Muller, M.~Bunka, S.~Haller, U.~Koster, V.~Groehn, P.~Bernhardt, N.~van~der
  Meulen, A.~Turler, R.~Schibli, Promising prospects for sc-44-/sc-47-based
  theragnostics: Application of sc-47 for radionuclide tumor therapy in mice,
  Journal of Nuclear Medicine 55~(10) (2014) 1658--1664.

\bibitem{12}
IAEA, Crp on nuclear data for charged-particle monitor reactions and medical
  isotope production (2012).

\bibitem{13}
F.~T\'ark\'anyi, F.~Szelecs\'enyi, S.~Tak\'acs, Determination of effective
  bombarding energies and fluxes using improved stacked-foil technique, Acta
  Radiologica, Supplementum 376 (1991) 72.

\bibitem{14}
NuDat, Nudat2 database (2.6) (2014).

\bibitem{15}
B.~Pritychenko, A.~Sonzogni, Q-value calculator (2003).

\bibitem{16}
H.~H. Andersen, J.~F. Ziegler, Hydrogen stopping powers and ranges in all
  elements. The stopping and ranges of ions in matter, Volume 3., The Stopping
  and ranges of ions in matter, Pergamon Press, New York, 1977.

\bibitem{17}
F.~T\'ark\'anyi, S.~Tak\'acs, K.~Gul, A.~Hermanne, M.~G. Mustafa, M.~Nortier,
  P.~Oblozinsky, S.~M. Qaim, B.~Scholten, Y.~N. Shubin, Z.~Youxiang, Beam
  monitor reactions (chapter 4). charged particle cross-section database for
  medical radioisotope production: diagnostic radioisotopes and monitor
  reactions., Tech. rep., IAEA (2001).

\bibitem{18}
{International-Bureau-of-Weights-and-Measures}, Guide to the expression of
  uncertainty in measurement, 1st Edition, International Organization for
  Standardization, Gen\'eve, Switzerland, 1993.

\bibitem{19}
A.~J. Koning, D.~Rochman, J.~Kopecky, J.~C. Sublet, E.~Bauge, S.~Hilaire,
  P.~Romain, B.~Morillon, H.~Duarte, S.~van~der Marck, S.~Pomp, H.~Sjostrand,
  R.~Forrest, H.~Henriksson, O.~Cabellos, G.~S., J.~Leppanen, H.~Leeb,
  A.~Plompen, R.~Mills, Tendl-2015: Talys-based evaluated nuclear data library,
  (2015).

\bibitem{20}
A.~J. Koning, D.~Rochman, Modern nuclear data evaluation with the talys code
  system, Nuclear Data Sheets 113 (2012) 2841.

\bibitem{21}
M.~Herman, R.~Capote, B.~V. Carlson, P.~Oblozinsky, M.~Sin, A.~Trkov,
  H.~Wienke, V.~Zerkin, Empire: Nuclear reaction model code system for data
  evaluation, Nuclear Data Sheets 108~(12) (2007) 2655--2715.

\bibitem{22}
M.~Herman, R.~Capote, M.~Sin, A.~Trkov, B.~Carlson, P.~Oblozinsky, C.~Mattoon,
  H.~Wienke, S.~Hoblit, Y.-S. Cho, V.~Plujko, V.~Zerkin, Nuclear reaction model
  code empire-3.2 (malta) (2012).

\bibitem{23}
S.~Tanaka, M.~Furukawa, Excitation functions for (p, n) reactions with
  titanium, vanadium, chromium, iron and nickel up to ep=14 mev, J. Phys. Soc.
  Japan 14 (1959) 1269--1275.

\bibitem{24}
R.~D. Albert, (p,n) cross cection and proton pptical-model, parameters in the 4
  to 5.5 mev energy region, Phys. Rev. 115 (1959) 925.

\bibitem{25}
B.~W. Shore, N.~S. Wall, J.~W. Irvine, Interactions of 7.5 mev protons with
  copper and vanadium, Phys. Rev. R. 123 (1961) 276.

\bibitem{26}
H.~Taketani, W.~P. Alford, (p, n) cross sections on ti47, v51, cr52, co59, and
  cu63 from 4 to 6.5 mev, Phys. Rev. 125~(1) (1962) 291--294.

\bibitem{27}
J.~Wing, J.~R. Huizenga, (p, n) cross sections of v51, cr52, cu63, cu65, ag107,
  ag109, cd111, cd114, and la139 from 5 to 10.5 mev, Phys. Rev. 128~(1) (1962)
  280.

\bibitem{28}
L.~F. Hansen, R.~D. Albert, Statistical theory predictions for 5- to 11-mev
  (p,n) and (p,p´) nuclear reactions in 51v, 59co, 63cu, 65cu, and 103rh,
  Physical Review 128 (1962) 291.

\bibitem{29}
G.~Albouy, M.~Gusakow, N.~Poffé, H.~Sergolle, L.~Valentin, Réaction (p,n) a
  moyenne énergie, Journal de Physique IV 23 (1962) 100.

\bibitem{30}
S.~Hontzeas, L.~Yaffe, Interaction of vanadium with protons of energies up to
  84 mev, Canadian Journal of Chemistry 41 (1963) 2194.

\bibitem{31}
R.~M. Humes, G.~F. Dell, W.~D. Ploughe, H.~J. Hausman, (p,n) cross sections at
  6.75 mev, Phys. Rev. 130 (1963) 1522.

\bibitem{32}
K.~K. Harrish, H.~A. Grench, R.~G. Johnson, F.~J. Vaughn, The v51(p,n)cr51
  reaction as a neutron source of known intensity, Nuclear Instruments \&
  Methods in Physics Research 33 (1965) 257.

\bibitem{33}
G.~Chodil, R.~C. Jopson, H.~Mark, C.~D. Swift, R.~G. Thomas, M.~K. Yates, (p,n)
  and (p,2n) cross sections on nine elements, Nuclear Physics A 93~(3) (1967)
  648--672.

\bibitem{34}
E.~Gadioli, A.~M. Grassi~Strini, G.~L. Bianco, G.~Strini, G.~Tagliaferri,
  Excitation functions of 51v, 56fe, 65cu(p, n) reactions between 10 and 45
  mev, Nuovo Cimento A 22 (1974) 547.

\bibitem{35}
J.~N. Barrandon, J.~L. Debrun, A.~Kohn, R.~H. Spear, Study of level of ti, v,
  cr, fe, ni, cu and zn by activation with protons whose energy is limited to
  20 mev, Nuclear Instruments \& Methods 127~(2) (1975) 269--278.

\bibitem{36}
M.~K. Mehta, S.~Kailas, K.~K. Sekharan, Total (p,n) reaction cross-section
  study on v-51 over incident energy-range 1.56 to 5.53 mev, Pramana 9~(4)
  (1977) 419--434.

\bibitem{37}
R.~Michel, G.~Brinkmann, H.~Weigel, W.~Herr, Measurement and hybrid-model
  analysis of proton-induced reactions with v, fe and co, Nuclear Physics A
  322~(1) (1979) 40--60.

\bibitem{38}
R.~Michel, G.~Brinkmann, On the depth-dependent production of radionuclides
  (44<=59) by solar protons in extraterrestrial matter, Journal of
  Radioanalytical Chemistry 59~(2) (1980) 467--510.

\bibitem{39}
J.~L. Zyskind, C.~A. Barnes, J.~M. Davidson, W.~A. Fowler, R.~E. Marrs, M.~H.
  Shapiro, Competition effects in proton-induced reactions on v-51, Nuclear
  Physics A 343~(2) (1980) 295--314.

\bibitem{40}
S.~Kailas, S.~K. Gupta, S.~S. Kerekatte, C.~V. Fernandes, V-51(p,n)cr-51
  reaction from ep 1.9 to 4.5 mev, Pramana 24~(4) (1985) 629--635.

\bibitem{41}
M.~A.~V. Bastos, J.~L.~Q. Debritto, U.~M. Vinagre, A.~G. Dasilva, A production
  method for cr-51 at iens cyclotron, Radiochimica Acta 50~(4) (1990) 189--191.

\bibitem{42}
V.~N. Levkovskii, The cross-sections of activation of nuclides of middle-range
  mass (A=40-100) by protons and alpha particles of middle range energies
  (E=10-50 MeV), Inter-Vesy, Moscow, 1991.

\bibitem{43}
P.~Jung, Cross sections for the production of helium and long-living
  radioactive isotopes by protons and deuterons, Tech. rep. (1991).

\bibitem{44}
M.~M. Musthafa, M.~K. Sharma, B.~P. Singh, R.~Prasad, Measurement and analysis
  of cross sections for (p,n) reactions in v-51 and in-113, Applied Radiation
  and Isotopes 62~(3) (2005) 419--428.

\bibitem{45}
G.~F. Dell, W.~D. Ploughe, H.~J. Hausman, Total reaction cross section in the
  mass range 45 to 65, Nuclear Physics 64 (1965) 513.

\bibitem{46}
R.~F. Carlson, A.~J. Cox, N.~E. Davison, R.~H. McCamis, W.~T.~H. van Oers, A
  study of proton total reaction cross sections for several medium mass nuclei
  between 20 and 48 mev, Bulletin American Physical Soc. 30~(1) (1985) 21.

\bibitem{47}
IAEA-NDS, Thin layer activation (tla) technique for wear measurement (2010).

\bibitem{48}
R.~Michel, F.~Peiffer, R.~Stuck, Measurement and hybrid model analysis of
  integral excitation-functions for proton-induced reactions on vanadium,
  manganese and cobalt up to 200 mev, Nuclear Physics A 441~(4) (1985)
  617--639.

\bibitem{49}
W.~Zhao, H.~Lu, W.~Yu, Excitation function of 51v(p, n)51cr up to 22 mev,
  Chinese Journal of Nuclear Physics 16~(1) (1994) 67.

\bibitem{50}
C.~G. Heininger, E.~O. Wiig, Production cross sections and yields of long lived
  44Τi from 100 mev proton bombardment of vanadium, Phys. Rev. 101 (1956)
  1074.

\bibitem{51}
S.~Tak\'acs, F.~T\'ark\'anyi, M.~Sonck, A.~Hermanne, Investigation of the
  mo-nat(p,x)tc-96mg nuclear reaction to monitor proton beams: New measurements
  and consequences on the earlier reported data, Nuclear Instruments Methods in
  Physics Research Section B-Beam Interactions with Materials and Atoms
  198~(3-4) (2002) 183--196.

\bibitem{52}
N.~G. Zaitseva, E.~Rurarz, M.~B. Tchikalov, M.~Vobecky, V.~A. Khalkin, L.~M.
  Popinenkova, Production cross sections and yields of long lived 44Τi from
  100 mev proton bombardment of vanadium, Radiochimica Acta 65~(3) (1994)
  157--160.

\bibitem{53}
M.~Bonardi, The contribution to nuclear data for biomedical radioisotope
  production from the milan cyclotron facility (1987).

\bibitem{54}
L.~Kanza-Kanza, M.~Cogneau, B.~Mahieu, D.~J. Apers, Simultaneous preparation of
  carrier-free cr-48, ti-45 and sc isotopes from cyclotron-irradiated vanadium
  targets, Radiochemical and Radioanalytical Letters 54~(1) (1982) 7--15.

\bibitem{55}
P.~P. Dmitriev, G.~A. Molin, Radioactive nuclide yields for thick target at 22
  mev proton energy, Vop. At. Nauki i Tekhn., Ser.Yadernye Konstanty 44~(5)
  (1981) 43.

\bibitem{56}
E.~Corniani, M.~Jech, T.~Wopelka, F.~Ditr\'oi, F.~Franek, A.~Pauschitz,
  High-resolution wear analysis of a ball-on-disc contact using low-activity
  radioactive isotopes, Proceedings of the Institution of Mechanical Engineers
  Part C-Journal of Mechanical Engineering Science 226~(C2) (2012) 319--326.

\bibitem{57}
F.~Ditr\'oi, F.~T\'ark\'anyi, S.~Tak\'acs, Wear measurement using radioactive
  tracer technique based on proton, deuteron and alpha-particle induced nuclear
  reactions on molybdenum, Nuclear Instruments \& Methods in Physics Research
  Section B-Beam Interactions with Materials and Atoms 290 (2012) 30--38.

\bibitem{58}
F.~T\'ark\'anyi, F.~Ditr\'oi, S.~Tak\'acs, A.~Hermanne, M.~Baba, A.~V.
  Ignatyuk, Investigation of activation cross-sections of deuteron induced
  reactions on vanadium up to 40 mev, Nuclear Instruments \& Methods in Physics
  Research Section B-Beam Interactions with Materials and Atoms 269~(15) (2011)
  1792--1800.

\end{thebibliography}







\end{document}